\documentclass[pdftex]{article}
\usepackage{color}
\usepackage[dvipdfmx]{graphicx}       
\begin{document}

\begin{center}{\Large Fitting State-space Model for Long-term Prediction \\[2mm]
of the Log-likelihood of Nonstationary Time Series Models}\\

\vspace{5mm}
{\large Genshiro Kitagawa}\\
Mathematics and Informatics Center, The University of Tokyo

\vspace{3mm}
{\today}

\end{center}

\noindent{\bf Abstract}
The goodness of the long-term prediction in the state-space model was evaluated 
using the squared long-term prediction error. 
In order to estimate the model parameters suitable for long-term prediction, 
we devised a modified log-likelihood corresponding to the long-term prediction error variance. 
Trend models and seasonally adjusted models with and without AR component are examined as examples.

\section{Introduction: State-Space Modeling of Time Series}

\subsection{State-Space Model and State Estimation}
Consider the state-space model of a univaraite time series $y_n$;
\begin{eqnarray}
 x_n &=& F_n x_{n-1} + G_n v_{n}, \hspace{1.1cm}\mbox{(system model)}\label{pr-1-1}\index{system model}\\
 y_n &=& H_n x_n + w_n, \hspace{1.78cm}\mbox{(observation model)}\label{pr-1-2}
\index{observation model}
\end{eqnarray}
\noindent
where $x_n$ is a $k$-dimensional state vector, $v_n$ is an $m$-dimensional system noise that follows a white noise with mean vector zero and variance-covariance matrix $Q_n$, and $w_n$ is the observation noise that follows a 1-dimensional Gaussian white noise with mean zero and the variance $R_n$. 
$F_n$, $G_n$, and $H_n$ are $k \times k$, $k \times m$, $k \times 1$ and $1 \times k$ matrices, respectively.
The initial state vector $x_0$ is assumed to follow the Gaussian distribution, $N(0,V_{0|0})$. 
Many linear models used in time series analysis such as the AR model, ARMA model and various nonstationary models such as the trend model and the seasonal adjustment model are expressible in terms of the state-space models (Anderson and Moore (1979), Kitagawa (2020)).

In this paper, we shall consider the problem of estimating the state $x_n$ at time $n$ based on the set of observations $Y_j =\{y_1,\cdots,y_j\}$. For $j<n$, $j=n$, and $j>n$, the state estimation problem is referred to as the prediction, filter and smoothing, respectively.
This state estimation problem is important in state-space modeling since many tasks such as one-step-ahead and multi-step-ahead prediction, interpolation, and likelihood computation for the time series can be systematically solved through the estimated state.

A generic approach to these state estimation problems is to obtain the conditional distribution\index{conditional distribution} $p(x_n|Y_j)$ of the state $x_n$ given the observations $Y_j$. 
Then, since the state-space model defined by (\ref{pr-1-1}) and (\ref{pr-1-2}) is a linear model, and moreover the noises $v_n$ and $w_n$, and the initial state $x_0$ follow normal distributions, all these conditional distributions become normal distributions. 
Therefore, to solve the problem of state estimation of the state-space model, it is sufficient to obtain the mean vectors $x_{n|j}$ and the variance-covariance matrices $V_{n|j}$ of the conditional distributions. 

For the linear state-space model, Kalman filter provides a computationally efficient recursive computational algorithm for state estimation (Kalman (1960), Anderson and Moore (1976)). 

\vskip2mm
{\bf One-step-ahead prediction}
%
\begin{eqnarray}
 x_{n|n-1} &=& F_n x_{n-1|n-1} \nonumber \\
 V_{n|n-1} &=& F_n V_{n-1|n-1} F_n^T + G_n Q_{n} G_n^T. \label{pr-3-2}
\end{eqnarray}
\indent
{\bf Filter} 
\begin{eqnarray}
 K_n &=& V_{n|n-1}H_n^T (H_n V_{n*n-1} H_n^T + R_n)^{-1} \nonumber \\
 x_{n|n} &=& x_{n|n-1} + K_n (y_n -H_n x_{n|n-1}) \label{pr-3-3} \\
 V_{n|n} &=& (I -K_n H_n)V_{n|n-1}. \nonumber
\end{eqnarray}

\subsection{Likelihood Computation and Parameter Estimation for Time Series Models}

Assume that the state-space representation for a time series model specified by a parameter $\theta$ is given. When the time series $y_1, \cdots, y_N$ of length $N$ is given, the $N$ dimensional joint density function of $y_1, \cdots, y_N$ specified by this time series model is denoted by $f_N(y_1, \cdots, y_N|\theta)$. Then, the likelihood\index{likelihood} of this model is defined by 
$ L(\theta ) = f_N( y_1,\cdots,y_N|\theta )$.
Using the conditional distribution of $y_n$ given the previous observations, the likelihood
of the time series model can be expressed as a product of one-dimentional conditional density functions:
\begin{equation}
 L(\theta ) = \prod_{n=1}^N g_n(y_n|y_1,\cdots,y_{n-1},\theta ) = \prod_{n=1}^N g_n(y_n|Y_{n-1},\theta ).
\end{equation}
Here, if we define $Y_0 =\emptyset$ (empty set), then $g_1(y_1|Y_0, \theta) \equiv f_1(y_1|\theta ) $. By taking the logarithm of $L(\theta)$, the log-likelihood of the model is obtained as
\begin{equation}
 \ell (\theta ) = \log L(\theta ) = \sum_{n=1}^N \log g_n(y_n|Y_{n-1},\theta ). \label{pr-6-1}
\end{equation}

Since $g_n(y_n|Y_{n-1},\theta)$ is the conditional distribution of $y_n$ given the observation $Y_{n-1}$ and it is, in fact, a normal distribution with mean $y_{n|n-1}$ and variance $d_{n|n-1}$, it can be expressed as (Kitagawa and Gersch (1996))
\begin{eqnarray}
g_n(y_n|Y_{n-1},\theta ) = 
 \left(2\pi d_{n|n-1}\right)^{-\frac{1}{2}}  \exp \left\{ -\frac{(y_n-y_{n|n-1})^2}{2 d_{n|n-1}} \right\}.
\end{eqnarray}
Here, from the observation model, (2), $y_{n|n-1}$ and $d_{n|n-1}$ are obtained by
\begin{eqnarray}
 y_{n|n-1} &=& H_{n}x_{n|n-1} \label{pr-5-1} \\
 d_{n|n-1} &=& H_{n}V_{n|n-1}H_{n}^T + R_{n}. \label{pr-5-2}
\end{eqnarray}

Therefore, by substituting this density function into (\ref{pr-6-1}), the log-likelihood of this state-space model is obtained as
\begin{eqnarray} 
 \ell (\theta ) =  -
\frac{1}{2} \biggl\{ N \log 2\pi + \sum_{n=1}^N \log d_{n|n-1} 
 + \sum_{n=1}^N \frac{(y_n-y_{n|n-1})^2}{d_{n|n-1}} \biggr\}. \label{pr-6-2} 
\end{eqnarray}

The maximum likelihood estiamtes of the parameters of the state-space model can be obtained by 
maximizing this log-likelihood function numerically.
However, for univariate time series, we can assume that $R=1$ (Kitagawa (2020)).
Actually, if $\tilde V_{n|n}$, $\tilde V_{n|n-1}$, $\tilde Q_{n}$, and $\tilde R$ are defined by
\begin{eqnarray}
 V_{n|n-1} = \sigma^2 \tilde V_{n|n-1}, \quad V_{n|n} = \sigma^2 \tilde V_{n|n}, \quad
   Q_{n} = \sigma^2 \tilde Q_{n}, \quad \tilde R = 1,
\end{eqnarray}
\noindent
then it can be seen that  the obtained Kalman gain $\tilde K_n$ is identical to $K_n$. Therefore, in the filtering step, we may use $\tilde V_{n|n}$ and $\tilde V_{n|n-1}$ instead of $V_{n|n}$ and $V_{n|n-1}$. Furthermore, it can be seen that the vectors $x_{n|n-1}$ and $x_{n|n}$ of the state do not change under these modifications. In summary, if $R_n$ is time-invariant and $R = \sigma^2$ is an unknown parameter, we may apply the Kalman filter by setting $R=1$. Since we then have $d_{n|n-1} = \sigma^2 \tilde d_{n|n-1}$ from (\ref{pr-6-2}), this yields
\begin{eqnarray}
 \ell (\theta ) = -\frac{1}{2}\left\{ N\log 2\pi \sigma^2 + \sum_{n=1}^N \log \tilde d_{n|n-1} + \frac{1}{\sigma^2} \sum_{n=1}^N \frac{(y_n-y_{n|n-1})^2}{ \tilde d_{n|n-1}} \right\}. \nonumber \\
\label{pr-6-3}
\end{eqnarray}

From the likelihood equation
\begin{equation}
 \frac{\partial \ell (\theta )}{\partial \sigma^2} = -\frac{1}{2} \left\{ \frac{N}{\sigma^2} 
-\frac{1}{(\sigma^2)^2}\sum_{n=1}^N \frac{(y_n-y_{n|n-1})^2}{ \tilde d_{n|n-1}} \right\} = 0,
\end{equation}
the maximum likelihood estimate\index{maximum likelihood estimate} of $\sigma^2$ is obtained by
\begin{equation}
 \hat \sigma^2 = \frac{1}{N} \sum_{n=1}^N \frac{(y_n-y_{n|n-1})^2}{ \tilde d_{n|n-1}}. \label{pr-6-4}
\end{equation}
Furthermore, denoting the parameters in $\theta$ except for the variance $\sigma^2$ by $\theta^{\ast}$, and substituting (\ref{pr-6-4}) into (\ref{pr-6-3}), we have an expression for the log-likelihood
\begin{equation}
 \ell (\theta^{\ast} ) = -\frac{1}{2}\left( N\log 2\pi \hat \sigma^2 + \sum_{n=1}^N \log \tilde d_{n|n-1} + N \right). \label{pr-6-5}
\end{equation}

\subsection{Parameter Estimation and Criterion for Increasing Horizon Prediction of the State}

For the state-space model, by repeating the one-step-ahead prediction step, we can perform increasing horizon prediction, that is, we can obtain $x_{n+j|n}$ and $V_{n+j|n}$ for $j=1, 2, \cdots p$.

\vskip2mm
{\bf The increasing horizon prediction} 

{}\qquad For $j=1,\cdots,p$, repeat
%
\begin{eqnarray}
 x_{n+j|n} &=& F_{n+j} x_{n+j-1|n} \nonumber \\
 V_{n+j|n} &=& F_{n+j} V_{n+j-1|n} F_{n+j}^T \: + \: G_{n+j} Q_{n+j} G_{n+j}^T. \label{pr-4-2}
\end{eqnarray}

The long-term prediction is considered by many authors such as Judd and Small (2000),
Sorjamaa et. al. (2007) and Xiong et. al. (2013).
In this paper, we evaluate the goodness of the long-term prediction by the difference between the
predictied value and the observed value
\begin{eqnarray}
  \hat\sigma^2_p = \frac{1}{N-p} \sum_{n=1}^{N-p} \varepsilon_{n+p|n}^2,
\end{eqnarray}
where $p$-step-ahead prediction error is defined by $\varepsilon_{n+p|n} = y_{n+p}-y_{n+p|n}$ and $y_{n+p|n}$ is defined by $y_{n+p|n} = Hx_{n+p|n}$.
We can also consider a modified log-likelihood for the long-term prediction defined by
\begin{eqnarray}
 \ell_p(\theta) = -\frac{1}{N-p} \left\{ (N-p)(\log 2\pi \hat\sigma^2_p  + 1) 
                + \sum_{n=1}^{N-p} \log d_{n+p|n} \right\} \label{EQ_long-term_presictive_likelihood}
\end{eqnarray}
where $d_{n+k|n}$ is obtained by $ d_{n+p|n} = H_{n+p} V_{n+p|n} H_{n+p}^T + R_{n+p}$.
Note that, different from the case of one-step-ahead prediction errors, 
the long-term prediction errors, $\varepsilon_{p+1|1},\ldots ,\varepsilon_{N|N-p} $
are not independent.

Given the predetermined prediction horizon $p$, the optimal value of the parameter vector $\theta$ for 
$p$-step-ahead prediction is obtained by maximizing this modified log-likelihood.

\section{Examples}
\subsection{Trend models}
\subsubsection{The secon order trend model}
We consider the second order trend model
\begin{eqnarray}
   y_n = T_n + w_n,
\end{eqnarray}
where $T_n$ is the trend component that follows the second order trend component model
$T_n = 2T_{n-1} - T_{n-2} + v_n$ and $w_n$ and $v_n$ are Gaussian white noise
$w_n \sim N(0,\sigma^2)$ and $v_n \sim N(0,\tau^2)$, respectively.
Note that this trend model can be expressed by a state-space model by
\begin{eqnarray}
  x_n = \left[ \begin{array}{l} T_n \\ T_{n-1} \end{array} \right],\quad
  F = \left[ \begin{array}{cc} 2 & -1 \\ 1 & 0 \end{array} \right],\quad
  G = \left[ \begin{array}{c}  1 \\ 0 \end{array} \right], \quad
  H = \left[ \begin{array}{cc} 1 & 0 \end{array} \right] .
\end{eqnarray}

\begin{figure}[tbp]
\begin{center}
\includegraphics[width=160mm,angle=0,clip=]{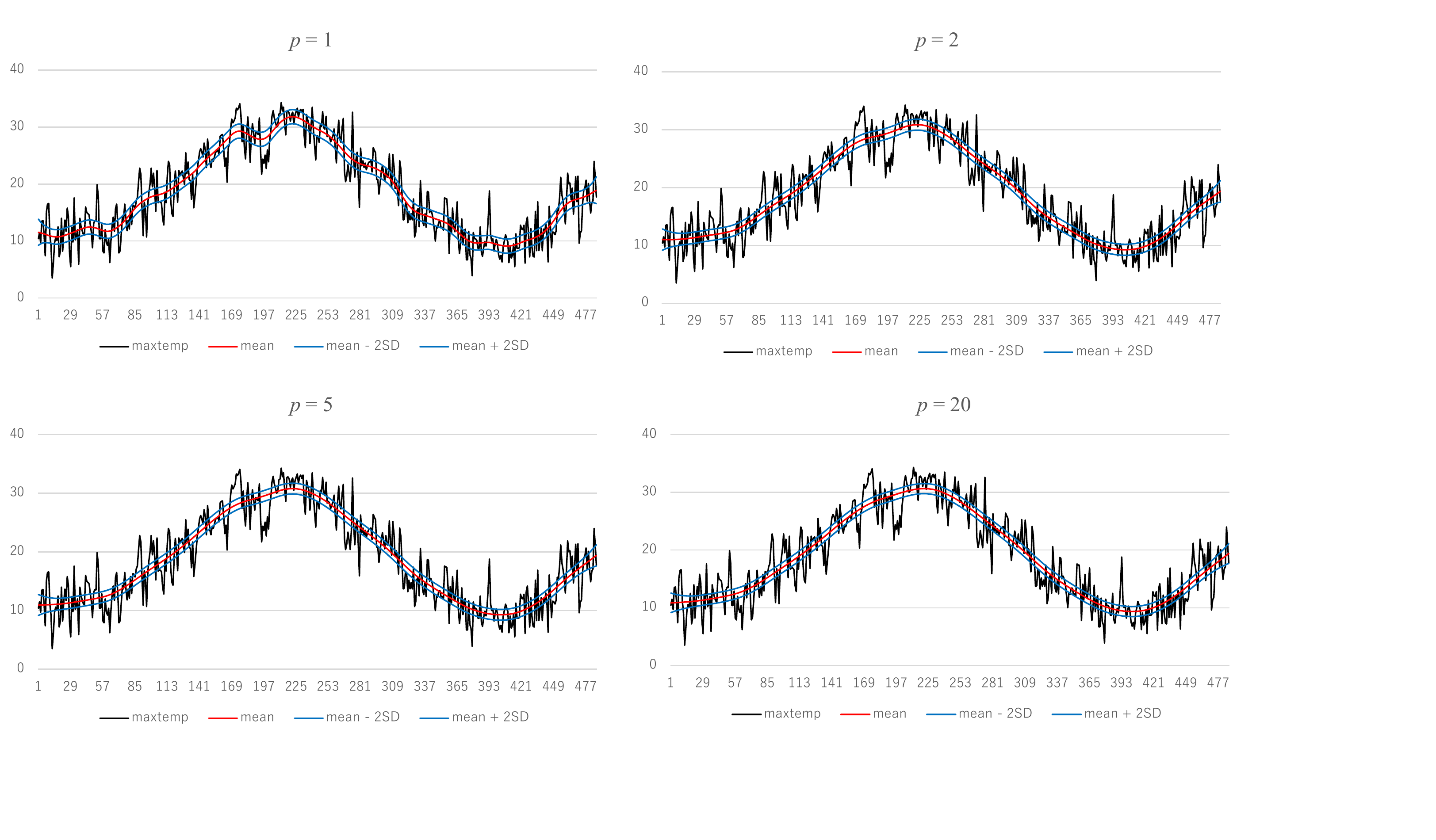}
\end{center}
\caption{The estimated trend of the maximum temperature data obtained by the second order trend models
estimated by the $p$-step-ahead prediction criterion, $p$=1, 2, 5 and 20. 
Each plot shows the data (black), the mean (red) , mean$\pm$ 2SD (blue) of the estimated trend components. }
\label{Fig_Trend model with order 2}
\end{figure}

Figure \ref{Fig_Trend model with order 2} shows the trend estimates of the
maxtemp data (daily maximum temperature data in Tokyo, N=486).
Smoothed estimate (red) and $\pm$2SD interval are shown.
Top left plot shows the estimates obtained by the ordinary maximum likelihood method,
i.e., by $p=1$.
On the other hand, three other plots show the results obtained by modified log-likelihood
criteria asumming $p=1,5$ and 20, respectively.
It can be seen that the trend estimate by $p=1$ is considerably variable compared with
other three estimates, and other three estimates obtained with $p>1$ resemble each other.

Table \ref{Tab_Prdictive_variance_trend_model_M1=2} shows the increasing horizon prediction error variances
$\hat\sigma_j^2$, $j=1,\ldots ,20$, for various parameter estimation criteria $\ell_p(\theta)$,
$p$=1,(1),6,(2),20.
The results for $p=1$ shown in the second column are the increasing horaizon prediction error
variances of the model obtained by the maximum-likelihood method.
In general, $(j,p)$-element of the table shows the $j$-step-ahead prediction error variance of
the model whose parameter was obtained by maximizing the modified log-likelihood for 
$p$-step-ahead prediction criterion (\ref{EQ_long-term_presictive_likelihood}).
Naturally, the one-step-ahead prediction error variance $\hat\sigma_1^2$ attains the smallest value
9.89 with $p=1$, but the long-term prediction error variances, 
i.e., for $j>1$, $\hat\sigma_j^2$  becomes the largest among $p=1,\ldots ,20$.
The table also shows that the $j$-step-ahead prediction error variance is the smallest at the criterion $p$.
For $p>1$, the increase of the long-term prediction error varaince is not so significant and
$\hat\sigma_j^2$ takes similar values for different $p$. The last row of the table shows the 
average of the long-term prediction error varainces, $\hat\sigma_j^2$, over $j=1,\ldots ,20$ for each $p$.

\begin{table}[tbp]
\begin{center}
\caption{Long-term prediction error variances of trend models with $m_1=2$ for various $p$.}\label{Tab_Prdictive_variance_trend_model_M1=2}

\vspace{2mm}\begin{small}
\begin{tabular}{c|cccccccccccccccc}
 & &&&&& $p$ \\
$j$  &   1&      2&      3&      4&      5&      6&      8&     10&     12&     14&     16&     18&     20\\
\hline
1	& {}\,9.89&	10.02&	10.05&	10.07&	10.07&	10.08&	10.13&	10.15&	10.16&	10.17&	10.20&	10.22&	10.20\\
2	&10.92&	10.52&	10.52&	10.53&	10.53&	10.53&	10.56&	10.58&	10.58&	10.59&	10.62&	10.64&	10.62\\
3	&11.93&	11.18&	11.16&	11.16&	11.16&	11.16&	11.18&	11.19&	11.20&	11.20&	11.22&	11.24&	11.22\\
4	&12.78&	11.60&	11.57&	11.55&	11.55&	11.55&	11.55&	11.55&	11.56&	11.56&	11.58&	11.59&	11.58\\
5	&13.62&	12.07&	12.02&	12.00&	12.00&	11.99&	11.97&	11.98&	11.98&	11.98&	12.00&	12.01&	12.00\\
6	&14.46&	12.58&	12.51&	12.48&	12.48&	12.46&	12.43&	12.43&	12.43&	12.43&	12.44&	12.45&	12.44\\
7	&15.82&	13.60&	13.51&	13.47&	13.47&	13.45&	13.41&	13.40&	13.40&	13.40&	13.41&	13.42&	13.41\\
8	&17.32&	14.52&	14.41&	14.35&	14.35&	14.33&	14.26&	14.25&	14.24&	14.24&	14.24&	14.24&	14.24\\
9	&18.58&	15.22&	15.09&	15.02&	15.02&	14.98&	14.90&	14.88&	14.88&	14.87&	14.86&	14.87&	14.86\\
10	&19.23&	15.39&	15.23&	15.15&	15.15&	15.11&	15.02&	15.00&	14.99&	14.99&	14.98&	14.98&	14.98\\
11	&20.40&	15.88&	15.70&	15.62&	15.61&	15.57&	15.46&	15.43&	15.42&	15.41&	15.40&	15.40&	15.40\\
12	&20.88&	15.99&	15.80&	15.71&	15.71&	15.66&	15.55&	15.53&	15.52&	15.51&	15.50&	15.51&	15.50\\
13	&22.18&	16.64&	16.43&	16.33&	16.33&	16.27&	16.15&	16.11&	16.11&	16.10&	16.08&	16.08&	16.08\\
14	&23.56&	17.28&	17.04&	16.93&	16.93&	16.87&	16.73&	16.69&	16.68&	16.67&	16.65&	16.65&	16.66\\
15	&25.22&	18.00&	17.72&	17.59&	17.59&	17.52&	17.35&	17.31&	17.30&	17.28&	17.26&	17.26&	17.26\\
16	&27.35&	18.79&	18.46&	18.30&	18.30&	18.22&	18.01&	17.95&	17.94&	17.92&	17.89&	17.87&	17.89\\
17	&30.28&	20.38&	20.01&	19.82&	19.82&	19.72&	19.48&	19.41&	19.40&	19.37&	19.33&	19.31&	19.33\\
18	&32.65&	21.51&	21.10&	20.89&	20.89&	20.78&	20.52&	20.43&	20.42&	20.39&	20.35&	20.32&	20.35\\
19	&33.84&	21.88&	21.46&	21.25&	21.25&	21.14&	20.88&	20.80&	20.79&	20.77&	20.72&	20.71&	20.73\\
20	&35.03&	22.49&	22.07&	21.87&	21.86&	21.76&	21.51&	21.44&	21.43&	21.41&	21.38&	21.37&	21.38\\
\hline
    &21.51& 16.09&  15.89&  15.80&  15.80&  15.75&  15.64&  15.61&  15.60&  15.59&  15.58&  15.59&  15.58
\end{tabular}
\end{small}
\end{center}
\end{table}

\begin{figure}[tbp]
\begin{center}
\includegraphics[width=100mm,angle=0,clip=]{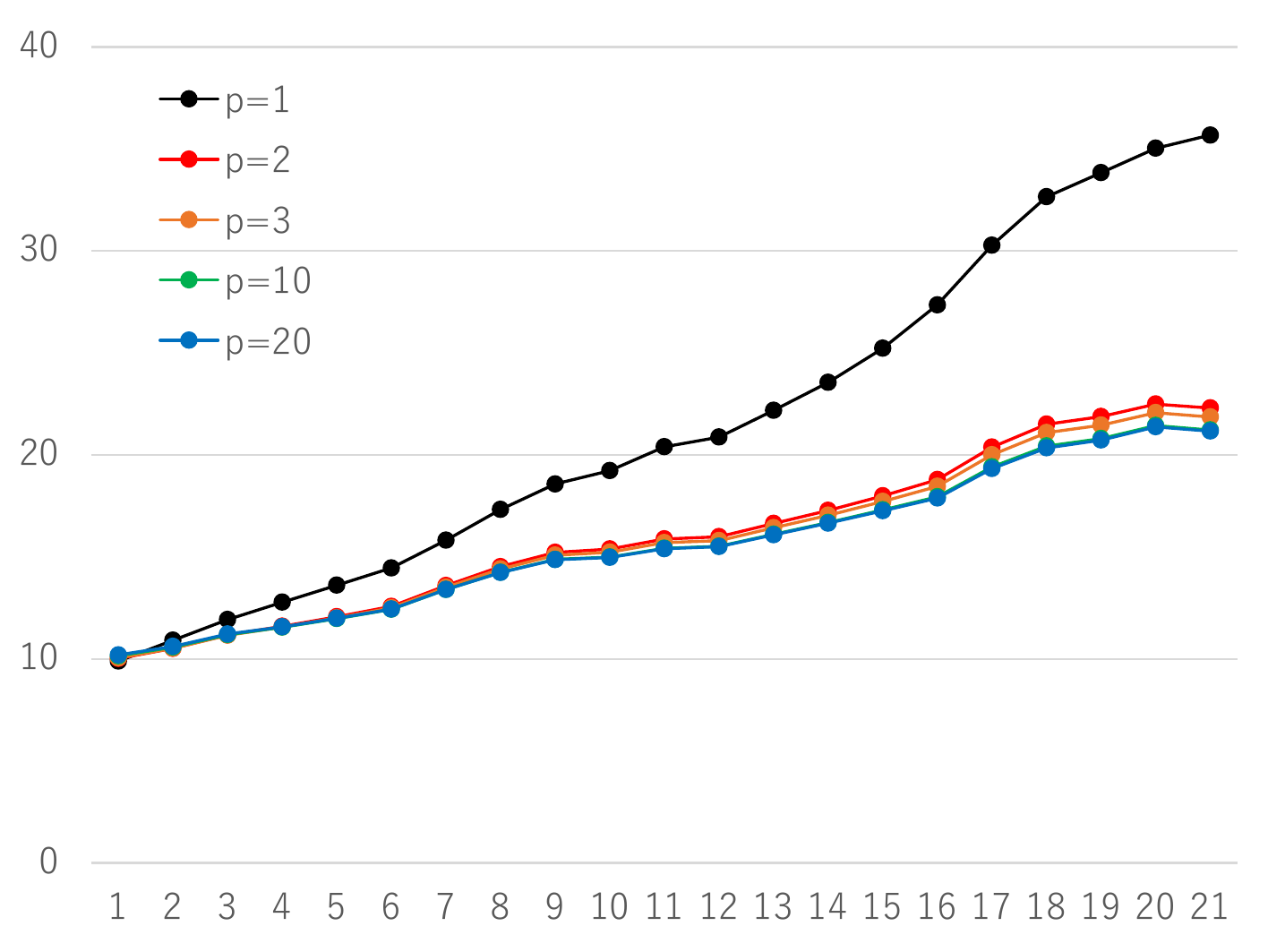}
\end{center}
\caption{The long-term prediction error variances of maximum temperature data by the second order trend model. 
Prediction lead time $p$=1, 2, 3, 10 and 20. }
\label{Fig_Error Varainces of trend model M=2}
\end{figure}

Figure \ref{Fig_Error Varainces of trend model M=2} shows the increase of the long-term prediction
error variance $\hat\sigma_j^2$, $j=1,\ldots ,20$ for $p=$1, 2, 3, 10 and 20.
We can see that the long-term prediction error variances obtained by $p=1$ are significantly 
larger than other cases, 
and there is almost no difference in the prediction error variances among $p=2$, 3, 10 and 20.

From the table and the figure, it can be concluded, 
at least for this data set, that the model with the maximum likelihood 
estimates of the patameter has the minimum one-step-ahead prediction error variance but has the largest
long-term prediction error varainces.
For this second order trend model $p=2,\ldots ,20$ yield a similar increasing horaizon prediction performances.
\subsubsection{The first order trend model}

\begin{figure}[tbp]
\begin{center}
\includegraphics[width=140mm,angle=0,clip=]{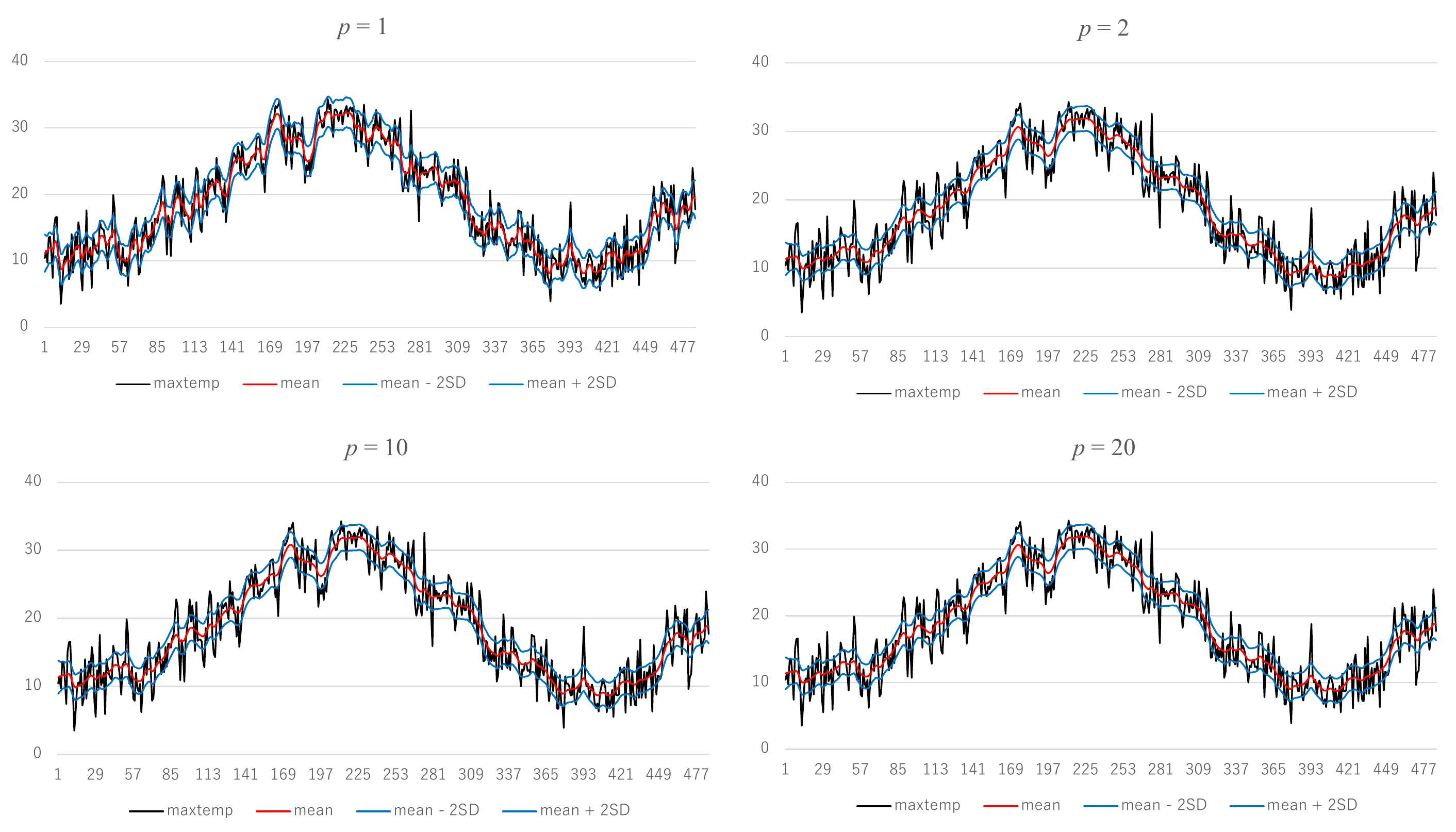}
\end{center}
\caption{The estimated trend of maximum temperature data by the first order trend model. 
Prediction lead time $p$=0, 1, 5 and 20. 
Each plot shows the data (black), the mean (red) , mean$\pm 2$SD (blue) of the estimated trend components. }
\label{Fig_Trend model with order 1}
\end{figure}

Figure \ref{Fig_Trend model with order 1} and Table \ref{Tab_Prdictive_variance_trend_model_M1=1}
show similar results for the first order trend model, 
\begin{eqnarray}
T_n = T_{n-1} + v_n,\qquad v_n \sim N(0,\tau^2).
\end{eqnarray}
As can be seen in the figure, 
in this case, different from the case of $k=2$, the estimated trends are wiggly and
the change of the estimated trends by the selection of the prediction horaizon $p$ 
are not so significant. 

From the table, it is seen that the one-step-ahead prediction error variances are
smaller than those of the second order trend model for entire $p$.
Also, the increase of the prediction error variance with the increase of the prediction 
horaizon is not so large as the second order trend model.
Further, it is interesting that the inreasing horaison predictive ability
is almost the same for all criterion parameter $p$.
The table also shows that the increasing horizon prediction performance is
quite similar for various choise of $p$.
Note that for the first order trend model, $j$-step-ahead prediction error
variance is obtained by
\begin{eqnarray}
 V_{n+j|n} = V_{n+j-1|n} + \tau^2.
\end{eqnarray}

Compared with the results of the second order trend model,
at least for the presend data, inspite of the wiggly trend estimate,
the first order trend model has slightly better prediction ability 
than the second order trend model.

\begin{table}[tbp]
\begin{center}
\caption{Long-term prediction error variances of trend models with $m_1=1$ for various $p$.}\label{Tab_Prdictive_variance_trend_model_M1=1}

\vspace{2mm}\begin{small}
\begin{tabular}{c|cccccccccccccccc}
    &   1&      2&      3&      4&      5&      6&      8&     10&     12&     14&     16&     18&     20\\
\hline
1&8.86 &9.06 &9.12 &9.17 &9.17 &9.12 &9.09 &9.06 &8.99 &8.90 &8.91 &9.07 &9.03 \\
2&10.34 &9.95 &9.95 &9.95 &9.95 &9.95 &9.94 &9.95 &9.97 &10.08 &10.07 &9.95 &9.95 \\
3&11.22 &10.68 &10.67 &10.66 &10.66 &10.67 &10.67 &10.68 &10.73 &10.89 &10.87 &10.68 &10.70 \\
4&12.01 &11.31 &11.28 &11.27 &11.27 &11.28 &11.30 &11.31 &11.39 &11.59 &11.57 &11.30 &11.34 \\
5&12.59 &11.82 &11.79 &11.78 &11.78 &11.79 &11.80 &11.82 &11.90 &12.12 &12.10 &11.81 &11.85 \\
6&12.88 &12.24 &12.22 &12.22 &12.22 &12.22 &12.22 &12.24 &12.29 &12.48 &12.45 &12.23 &12.25 \\
7&13.18 &12.67 &12.67 &12.68 &12.68 &12.67 &12.67 &12.67 &12.71 &12.85 &12.83 &12.67 &12.68 \\
8&13.88 &13.30 &13.30 &13.30 &13.30 &13.30 &13.30 &13.30 &13.35 &13.51 &13.49 &13.30 &13.32 \\
9&14.39 &13.79 &13.78 &13.79 &13.79 &13.78 &13.78 &13.79 &13.83 &14.00 &13.98 &13.78 &13.80 \\
10&14.70 &14.14 &14.14 &14.16 &14.16 &14.14 &14.14 &14.14 &14.17 &14.32 &14.30 &14.14 &14.15 \\
11&15.30 &14.67 &14.67 &14.68 &14.68 &14.67 &14.66 &14.67 &14.70 &14.87 &14.85 &14.66 &14.67 \\
12&15.30 &14.92 &14.96 &14.99 &14.99 &14.95 &14.94 &14.92 &14.91 &14.99 &14.98 &14.93 &14.91 \\
13&15.65 &15.42 &15.47 &15.51 &15.51 &15.46 &15.44 &15.42 &15.38 &15.41 &15.40 &15.43 &15.40 \\
14&15.87 &15.84 &15.90 &15.96 &15.96 &15.90 &15.87 &15.84 &15.77 &15.74 &15.74 &15.85 &15.81 \\
15&16.39 &16.48 &16.55 &16.61 &16.61 &16.55 &16.51 &16.48 &16.40 &16.34 &16.34 &16.49 &16.44 \\
16&17.52 &17.48 &17.53 &17.58 &17.58 &17.53 &17.50 &17.48 &17.43 &17.42 &17.42 &17.49 &17.46 \\
17&18.81 &18.52 &18.55 &18.58 &18.57 &18.54 &18.53 &18.52 &18.52 &18.59 &18.58 &18.52 &18.51 \\
18&19.99 &19.43 &19.43 &19.45 &19.45 &19.43 &19.43 &19.43 &19.47 &19.62 &19.61 &19.43 &19.44 \\
19&20.74 &20.11 &20.11 &20.13 &20.13 &20.11 &20.11 &20.11 &20.15 &20.32 &20.30 &20.11 &20.12 \\
20&20.97 &20.48 &20.51 &20.54 &20.54 &20.51 &20.49 &20.48 &20.49 &20.62 &20.60 &20.49 &20.48 \\
21&21.46 &21.00 &21.03 &21.07 &21.07 &21.03 &21.02 &21.00 &21.00 &21.11 &21.10 &21.01 &21.00 \\
\hline
    &15.34& 14.92&  14.93&  14.96&  14.96&  14.93&  14.92&  14.92&  14.93&  15.04&  15.02&  14.92&  14.92
\end{tabular}
\end{small}
\end{center}
\end{table}

\newpage
\subsection{Seasonal adjustment model}

\subsubsection{Standard seassonal adjustment model}
We consider here the seasonal adjustment model for the blsallhood data
(Buleau of Labor Statistics, all food data, N=156, Kitagawa (2020)),
\begin{eqnarray}
 y_n = T_n + S_n + w_n,
\end{eqnarray}
where $T_n$ and $S_n$ are the trend and the seasonal components that follow
\begin{eqnarray}
 T_n &=& 2T_{n-1} - T{n-2} + u_n \nonumber \\
 S_n &=& -(S_{n-1}+\cdots + S_{n-11}) + v_n.
\end{eqnarray}

\begin{figure}[tbp]
\begin{center}
\includegraphics[width=140mm,angle=0,clip=]{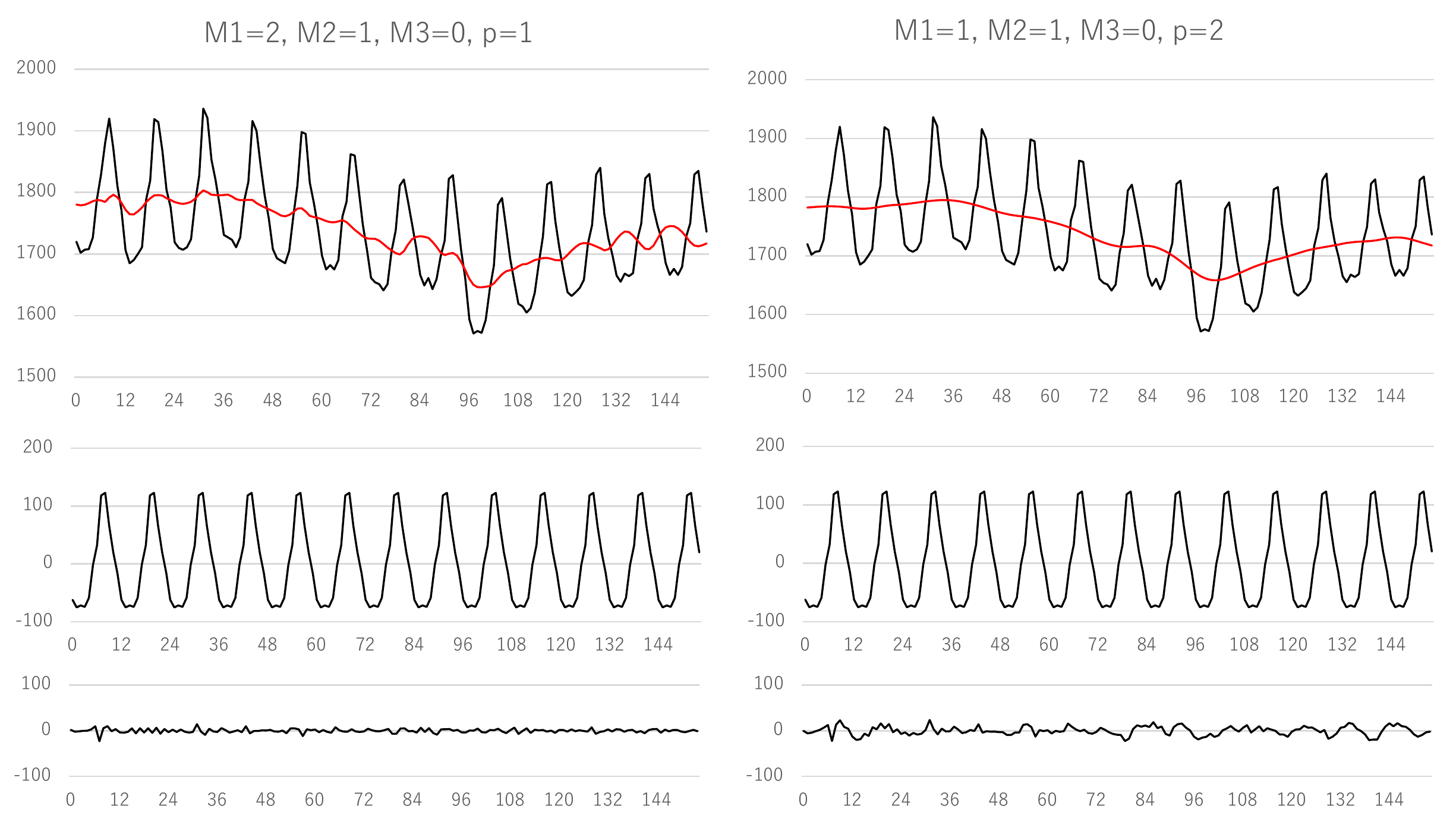}
\end{center}
\caption{The seasonal adjustment of blsallfood data with $m_1=2$ and $m_2=1$. 
Prediction lead time $p$=0 and 1. 
Top plot shows the data (black) and the mean of the trend(red), the second plot the seasonal component and the bottom plot shows the noise component. }
\label{Fig_SA model with order 2_1}
\end{figure}

\begin{figure}[tbp]
\begin{center}
\includegraphics[width=140mm,angle=0,clip=]{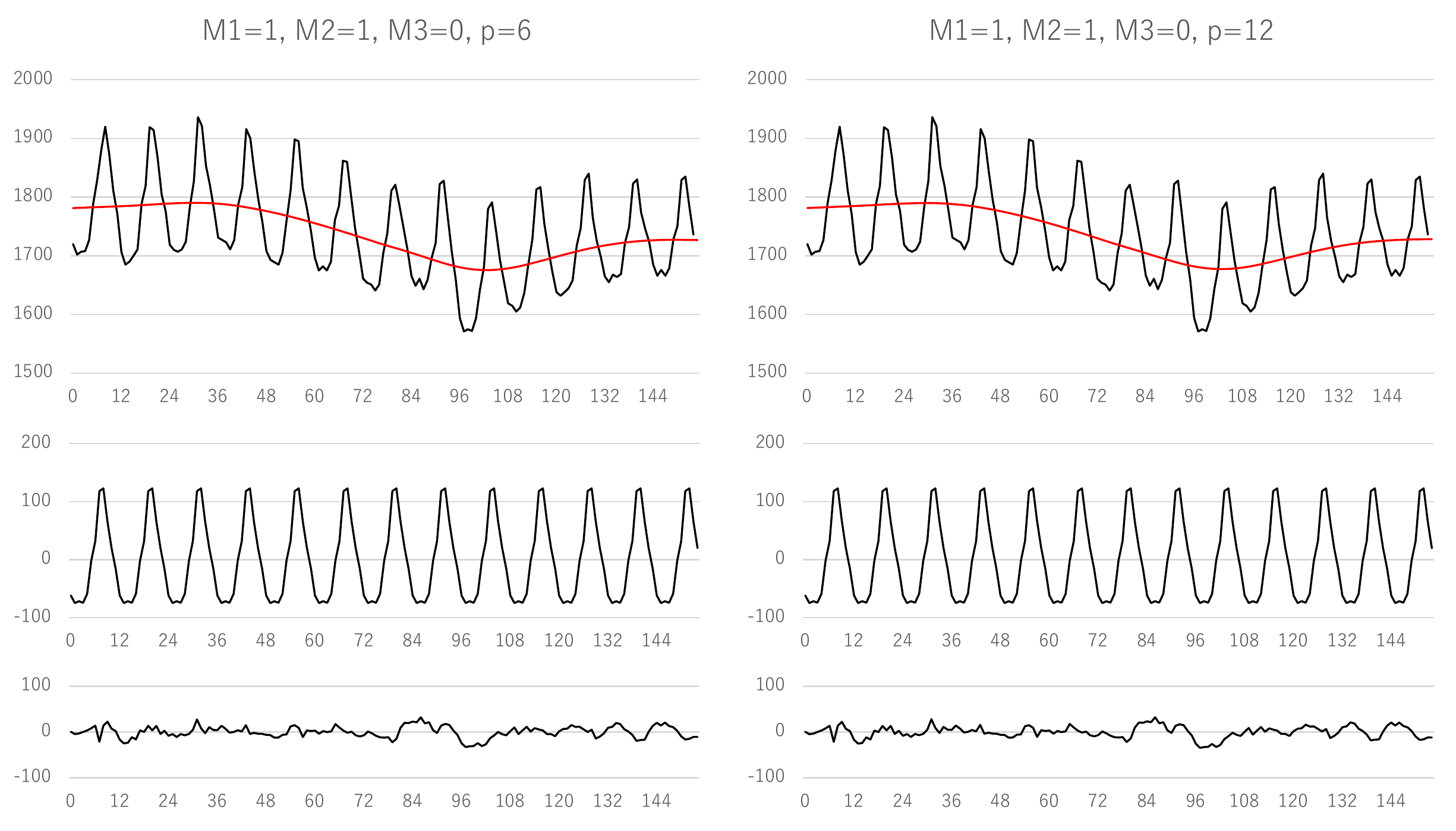}
\end{center}
\caption{The seasonal adjustment of blsallfood data with $m_1=2$ and $m_2=1$. 
Prediction lead time $p$=6 and 12. 
Top plot shows the data (black) and the mean of the trend(red), the second plot the seasonal component and the bottom plot shows the noise component. }
\label{Fig_SA model with order 2_2}
\end{figure}

Figure \ref{Fig_SA model with order 2_1} shows the decomposition of the data
into trend, seasonal component and the observation noise. The left plot
shows the case of the maximum likelihood estimate, i.e., obtained by $p=1$.
The right plot is the case of $p=2$.
It can be seen that compared with the standard results obtained by
the maximum likelihood estimates, the estimated trend by $p$=2 is smoother.
Similarly, Figure \ref{Fig_SA model with order 2_2} show the cases of
$p$=6 and 12.
In these cases, the trend become further smoother. No significan difference
is seen between the results of $p$=6 and 12.
The seasonal components are almost the same for all cases, and the observation
noise components of $p=6$ and $p=12$ are almost identical.

Table \ref{Tab_Prdictive_variance_TSmodel_M1=2} shows the increasing horaizon prediction error variances 
of the seasonal adjustment model obtained with $p=1,\ldots , 24$.
In this case, the one-step-ahead prediction error variance $\hat\sigma^2_1$ 
of the maximum likelihood estiamte, obtained by $p=1$, 
is 133 and is significantly smaller than other $p$.
This indicates that the maximum likelihood estimate has siginificantly better
one-step-ahead prediction performance than the other $p$.
However, the increase of the variance for increasing prediction horaizon $j$ is significant and
takes the largest long-term prediction variances for $j\geq 3$ among all prediction horizon $p=1,\ldots ,24$.
For $p\geq 6$ the prediction error varainces take similar values for entire $j$.
The last row of the table shows the average of the $\hat\sigma^2_j$ over $j=1,\ldots ,24$ for $j=1,\ldots 20$.
This averaged prediction error variance is significantly large for $p$=1 and
takes similar values for $p\geq 6$.
Note that this table suggests that if the increasing horaizon prediction is necessary
it is very likely that we can obtain a good prediction performance by taking $p$=6 or larger.

\begin{table}[tbp]
\begin{center}
\caption{Long-term prediction error variances of seasonal adjustment model with
$m_1=2$, $m_2=1$  for various $p$.}\label{Tab_Prdictive_variance_TSmodel_M1=2}

\vspace{2mm}
\tabcolsep=1mm
\begin{small}
\begin{tabular}{c|cccccccccccccccccc}
    &&&&&&& $p$ \\
$j$ &1&2&3&4&5&6&8&10&12&14&16&18&20&22&24\\
\hline
1 &	133 &	176 &	217 &	229 &	237 &	240 &	243 &	246 &	253 &	256 &	254 &	250 &	247 &	246 &	246 \\
2 &	311 &	281 &	293 &	300 &	306 &	308 &	311 &	314 &	319 &	322 &	320 &	317 &	314 &	314 &	314 \\
3 &	585 &	401 &	370 &	373 &	376 &	378 &	379 &	381 &	386 &	388 &	387 &	384 &	382 &	381 &	381 \\
4 &	936 &	525 &	445 &	442 &	443 &	444 &	445 &	446 &	449 &	451 &	450 &	448 &	446 &	446 &	446 \\
5 &	1359 &	646 &	516 &	507 &	506 &	506 &	507 &	507 &	510 &	511 &	511 &	509 &	508 &	507 &	507 \\
6 &	1786 &	754 &	578 &	565 &	563 &	562 &	562 &	563 &	565 &	566 &	566 &	564 &	563 &	563 &	563 \\
7 &	2324 &	863 &	641 &	624 &	620 &	619 &	619 &	619 &	621 &	622 &	622 &	620 &	619 &	619 &	619 \\
8 &	2797 &	956 &	700 &	679 &	674 &	673 &	673 &	673 &	675 &	676 &	676 &	674 &	673 &	673 &	673 \\
9 &	3183 &	1053 &	763 &	739 &	733 &	732 &	731 &	731 &	733 &	734 &	734 &	732 &	731 &	731 &	731 \\
10&	3582 &	1194 &	846 &	816 &	808 &	806 &	805 &	804 &	806 &	807 &	806 &	805 &	804 &	804 &	804 \\
11&	4010 &	1381 &	944 &	905 &	894 &	891 &	889 &	888 &	888 &	889 &	888 &	888 &	888 &	888 &	888 \\
12&	4520 &	1624 &	1062 &	1012 &	996 &	991 &	988 &	986 &	985 &	985 &	985 &	985 &	986 &	986 &	986 \\
13&	5317 &	1917 &	1203 &	1141 &	1122 &	1116 &	1112 &	1109 &	1106 &	1106 &	1106 &	1107 &	1108 &	1109 &	1109 \\
14&	6402 &	2179 &	1324 &	1252 &	1229 &	1222 &	1217 &	1213 &	1209 &	1209 &	1209 &	1210 &	1212 &	1213 &	1213 \\
15&	7515 &	2381 &	1425 &	1347 &	1323 &	1315 &	1310 &	1306 &	1302 &	1301 &	1302 &	1303 &	1305 &	1306 &	1306 \\
16&	8530 &	2548 &	1517 &	1436 &	1411 &	1403 &	1398 &	1394 &	1391 &	1391 &	1391 &	1392 &	1394 &	1394 &	1394 \\
17&	9296 &	2695 &	1607 &	1526 &	1502 &	1494 &	1489 &	1486 &	1484 &	1484 &	1484 &	1484 &	1485 &	1486 &	1486 \\
18&	10184 &	2844 &	1703 &	1621 &	1597 &	1590 &	1586 &	1583 &	1582 &	1584 &	1583 &	1582 &	1583 &	1583 &	1583 \\
19&	11034 &	2990 &	1809 &	1727 &	1704 &	1697 &	1694 &	1692 &	1692 &	1694 &	1693 &	1691 &	1691 &	1692 &	1692 \\
20&	11893 &	3152 &	1933 &	1849 &	1827 &	1821 &	1817 &	1816 &	1818 &	1821 &	1820 &	1817 &	1816 &	1816 &	1816 \\
21&	12633 &	3347 &	2080 &	1993 &	1969 &	1963 &	1960 &	1959 &	1962 &	1966 &	1964 &	1960 &	1959 &	1959 &	1959 \\
22&	13210 &	3612 &	2262 &	2166 &	2140 &	2133 &	2129 &	2128 &	2131 &	2134 &	2133 &	2129 &	2128 &	2128 &	2128 \\
23&	14240 &	3978 &	2482 &	2371 &	2339 &	2330 &	2326 &	2323 &	2324 &	2328 &	2326 &	2323 &	2323 &	2323 &	2323 \\
24&	15605 &	4383 &	2721 &	2592 &	2554 &	2543 &	2537 &	2533 &	2532 &	2534 &	2533 &	2531 &	2532 &	2533 &	2533 \\
\hline
  &	6308 &	1912 &	1227 &	1176 &	1161 &	1157 &	1155 &	1154 &	1155 &	1157 &	1156 &	1154 &	1154 &	1154 &	1154 
\end{tabular}
\end{small}
\end{center}
\end{table}

\begin{figure}[tbp]
\begin{center}
\includegraphics[width=120mm,angle=0,clip=]{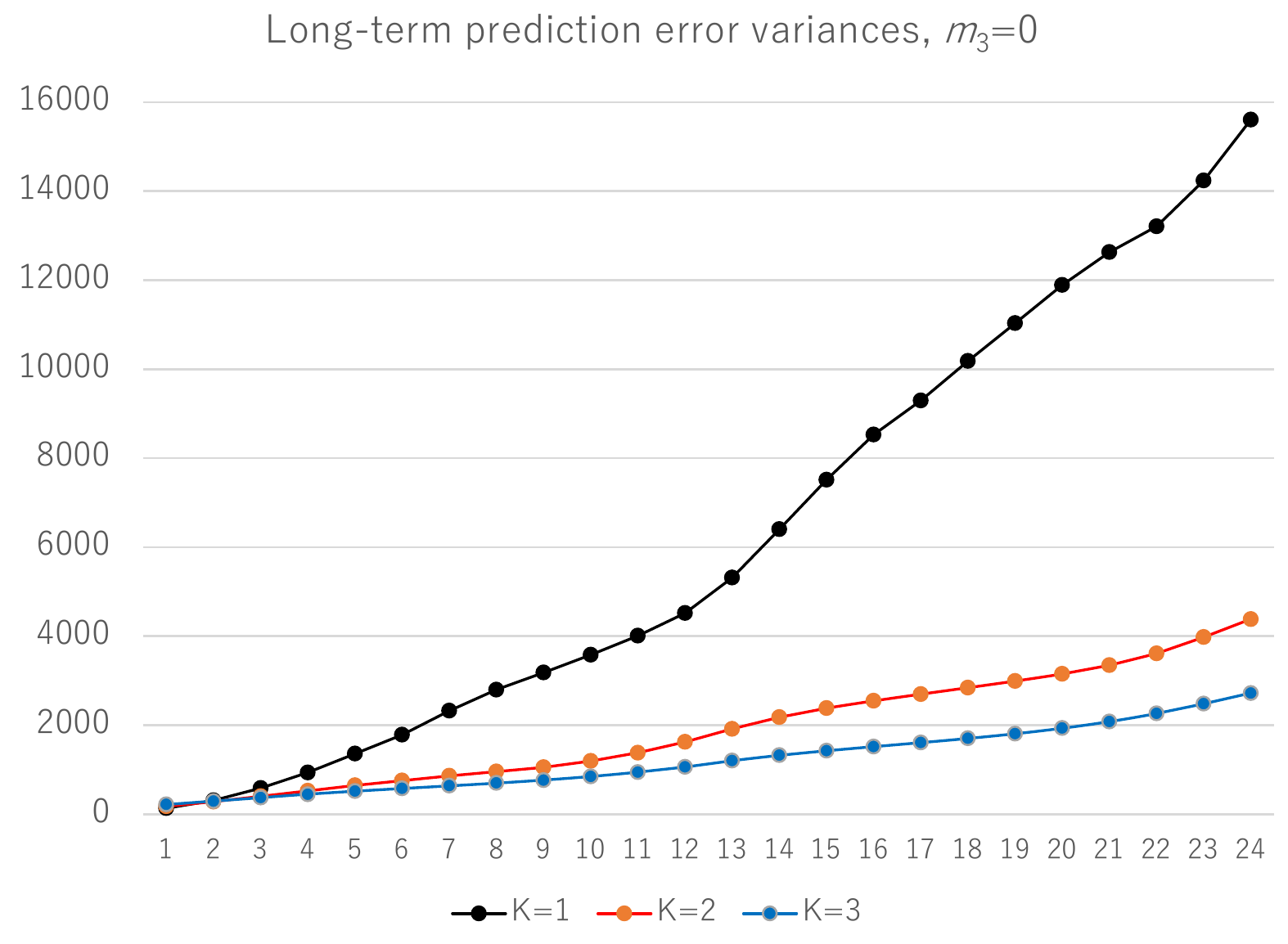}
\end{center}
\caption{The long-term prediction varainces of blsallfood data for increasing prediction horaizon (i=1,\ldots 20)
by the seasonal adjustment model with $m_1=2$, $m_2=1$ and $m_3=0$. 
$p$=1, 2 and 3.  }
\label{Fig_pred_error_variances}
\end{figure}

Figure \ref{Fig_pred_error_variances} shows the changes of increasing horizon prediction
of the seasonal adjustment model for $p=$1, 2 and 3.
The curves for $p \geq 4$ are visually indistiguishable from the curve for $p=3$.
From this figure, we can see that by using a $p$ greater than 2, 
the long-term prediction error variance can be reduced, instead of increasing the one-step-ahead
prediction error variance $\sigma^2_1$.

We also considered the increasing-horizon prediction performance of the
seasonal adjustment model with the first order trend model
\begin{eqnarray}
T_n = T_{n-1} + u_n,\qquad u_n \sim N(0,\tau^2 ).
\end{eqnarray}
Unlike the previous case of the seasonal adjustment with the second order trend model,
similarly to the case of first order trend model, the increasing horaizon prediction results are
almost the same for entire $p=1,\ldots 24$.
However,  for $p=14$ and 16, the prediction error variances are larger than other cases.
It is probable that the problem of optimization cased this anomalous phenomenon.

\begin{figure}[tbp]
\begin{center}
\includegraphics[width=140mm,angle=0,clip=]{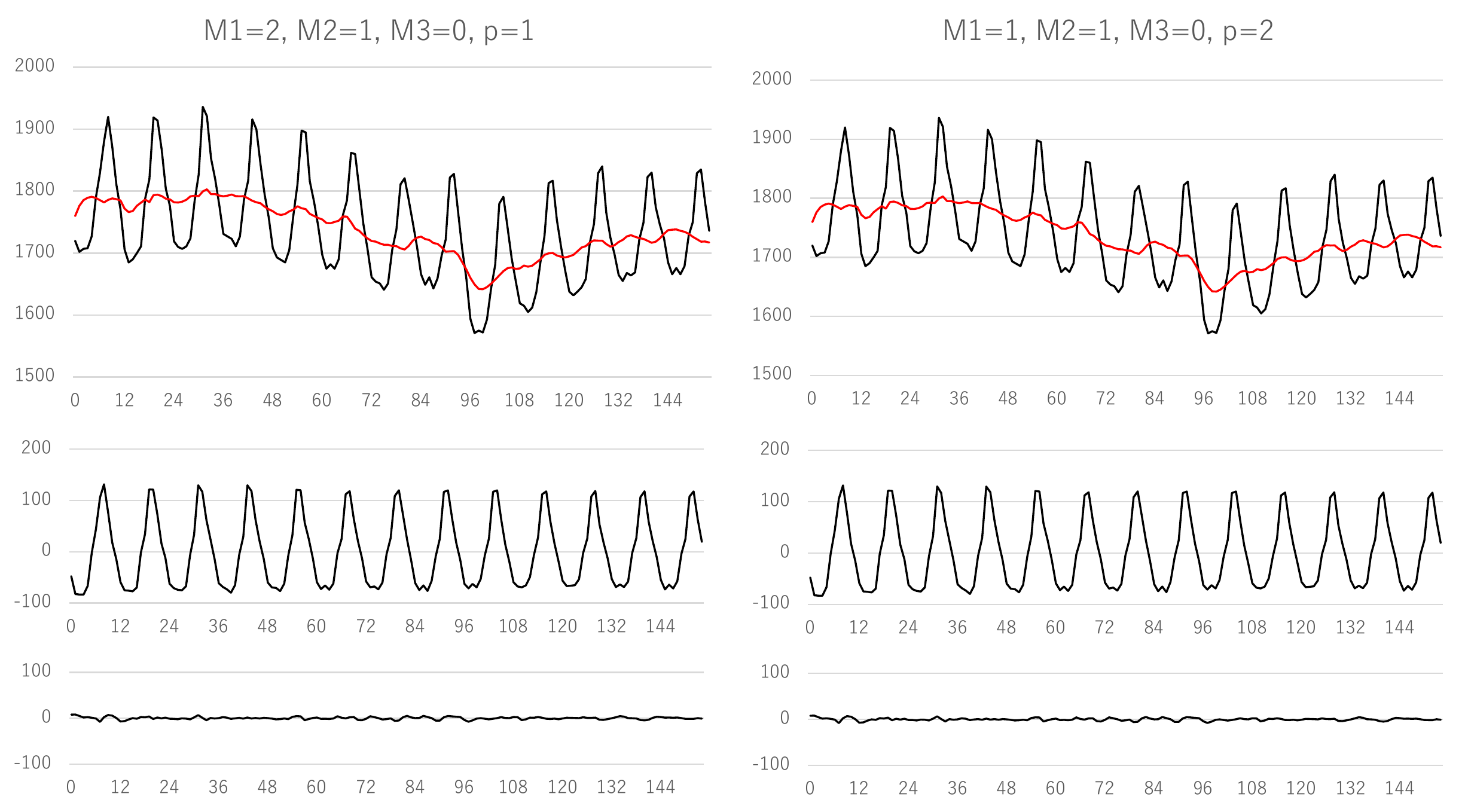}
\end{center}
\caption{The seasonal adjustment with $m_1=1$ and $m_2=1$. 
Prediction lead time $p$=0 and 1. 
Top plot shows the data (black) and the mean of the trend(red), the second plot the seasonal component and the botomn plot shows the noise component. }
\label{Fig_SA model with order 1_1}
\end{figure}

\begin{figure}[tbp]
\begin{center}
\includegraphics[width=140mm,angle=0,clip=]{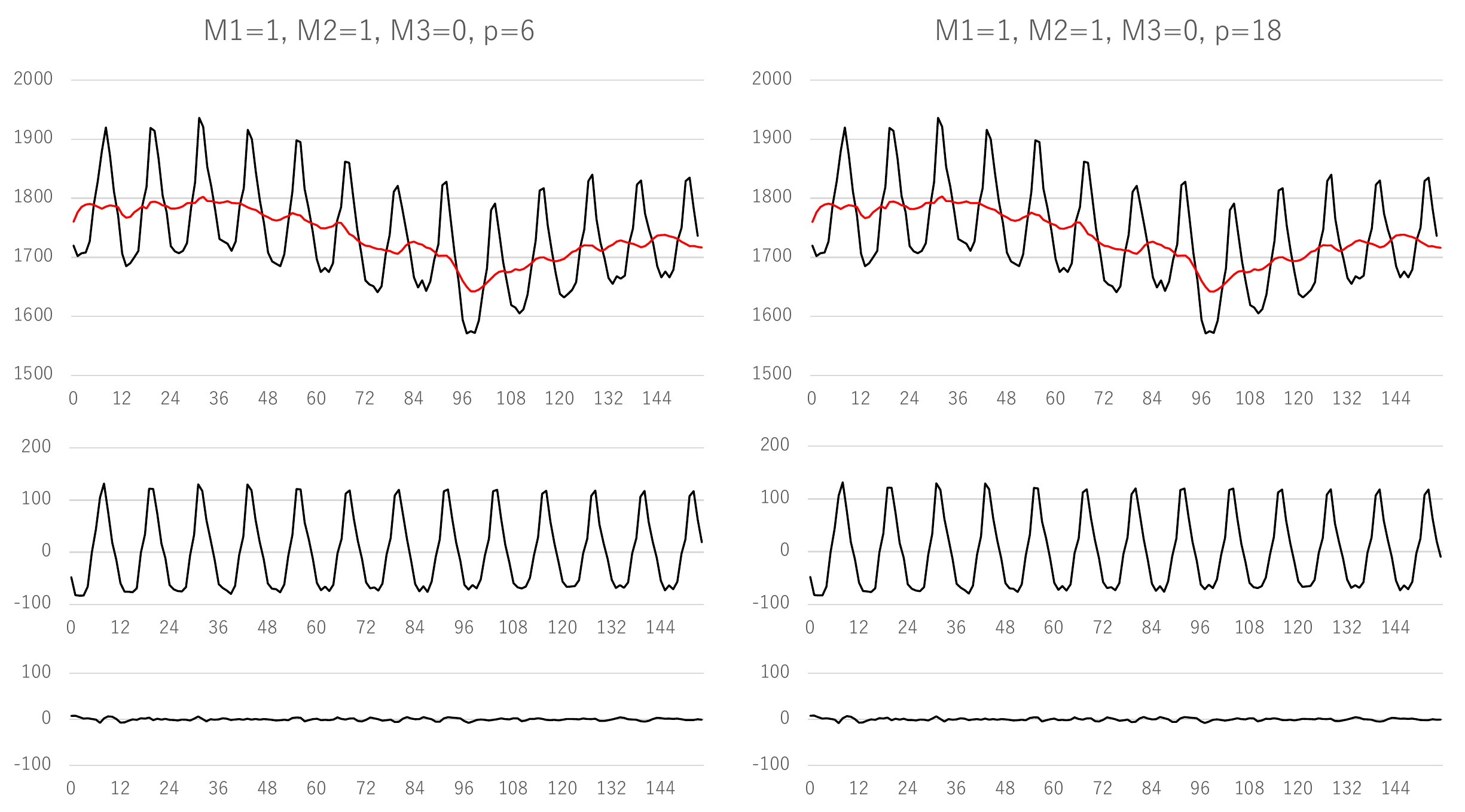}
\end{center}
\caption{The seasonal adjustment model $m_1=1$ and $m_2=1$. 
Prediction lead time $p$=6 and 12. 
Top plot shows the data (black) and the mean of the trend(red), the second plot the seasonal component and the botomn plot shows the noise component. }
\label{Fig_SA model with order 1_2}
\end{figure}

The figures and table for this case are shown in the Appendix as 
Figures \ref{Fig_pred_error_variances_TS_M1=1}, \ref{Fig_SA model with order 1_1}, 
\ref{Fig_SA model with order 1_2}, and Table \ref{Tab_Prdictive_variance_TSmodel_M1=2}.

\newpage
\subsubsection{Seasonal adjustment model with AR component}

Table \ref{Tab_Prdictive_variance_TSmodel_M1=2_M3=2} shows the long-term prediction error variances
when we used the seasonal adjustment model with stationary AR component (Kitagawa and Gersch (1974),
Kitagawa (2020)):
\begin{eqnarray}
   y_n = T_n + S_n + p_n + w_n,
\end{eqnarray}
where $p_n$ is the stationary AR components that follows an AR model with order $m_3$,
\begin{eqnarray}
   p_n = \sum_{j=1}^{m_3} a_j y_{n-j} + z_n,
\end{eqnarray}
$z_n \sim N(0,\tau^2_3)$.
Compared with the Table \ref{Tab_Prdictive_variance_TSmodel_M1=2} for the standard seasonal
adjustment model without AR component, the one-step-ahead prediction error variance is
smaller and the increase of the long-term prediction error variance 
for large $j$ is moderate, and no noticable changes are seen by the change 
of the prediction horaizon $p$.
The bottom row of the table shows that the minimum of the averaged prediction-error-variance
is attained at $p=8 \sim 20$.
It can be seen that the averaged prediction error variaces are almost the same for
$p=1 \sim 6$, and then takes the smallest value at $p=8 \sim 20$, and takes slightly
larger values for $p=22$ and 24.
The table also shows that, compared with the ordinary seasonal adjustment model, 
the seasonal adjustment model with AR component has high prediction performance
over entire prediction horizon, $j=1, 2, \ldots ,24$.

\begin{table}[tbp]
\begin{center}
\caption{Long-term prediction error variances of seasonal adjustment model with
$m_1=2$, $m_2=1$ and $m_3=2$ for various $p$.}\label{Tab_Prdictive_variance_TSmodel_M1=2_M3=2}

\vspace{2mm}
\tabcolsep=1mm
\begin{small}
\begin{tabular}{c|cccccccccccccccccc}
    &&&&&&& $p$ \\
$j$ &1&2&3&4&5&6&8&10&12&14&16&18&20&22&24\\
\hline
1 &	102 &	102 &	102 &	102 &	102 &	102 &	184 &	184 &	184 &	184 &	184 &	184 &	184 &	184 &	184 \\
2 &	187 &	187 &	187 &	187 &	187 &	188 &	209 &	209 &	209 &	209 &	209 &	209 &	209 &	210 &	210 \\
3 &	281 &	281 &	281 &	281 &	281 &	282 &	336 &	336 &	336 &	336 &	336 &	336 &	336 &	336 &	336 \\
4 &	369 &	369 &	369 &	369 &	370 &	370 &	368 &	368 &	368 &	368 &	368 &	368 &	368 &	370 &	370 \\
5 &	455 &	455 &	455 &	455 &	454 &	454 &	457 &	457 &	457 &	457 &	457 &	457 &	457 &	459 &	459 \\
6 &	517 &	517 &	517 &	516 &	516 &	516 &	484 &	484 &	484 &	484 &	484 &	484 &	484 &	488 &	488 \\
7 &	590 &	590 &	590 &	589 &	587 &	586 &	547 &	547 &	547 &	547 &	547 &	547 &	547 &	551 &	551 \\
8 &	643 &	644 &	643 &	643 &	639 &	637 &	569 &	569 &	569 &	569 &	569 &	569 &	569 &	574 &	574 \\
9 &	682 &	683 &	682 &	682 &	677 &	675 &	622 &	622 &	622 &	622 &	622 &	622 &	622 &	627 &	627 \\
10&	734 &	734 &	734 &	733 &	730 &	727 &	658 &	658 &	658 &	658 &	658 &	658 &	658 &	666 &	666 \\
11&	793 &	793 &	792 &	792 &	790 &	788 &	740 &	740 &	740 &	740 &	740 &	740 &	740 &	746 &	746 \\
12&	878 &	879 &	878 &	878 &	878 &	877 &	805 &	805 &	805 &	805 &	805 &	805 &	805 &	814 &	814 \\
13&	1002 &	1003 &	1002 &	1002 &	1002 &	1002 &	932 &	932 &	932 &	932 &	932 &	932 &	932 &	941 &	941 \\
14&	1133 &	1134 &	1133 &	1133 &	1131 &	1131 &	1003 &	1003 &	1003 &	1003 &	1003 &	1003 &	1003 &	1016 &	1016 \\
15&	1247 &	1248 &	1247 &	1247 &	1243 &	1243 &	1088 &	1088 &	1088 &	1088 &	1088 &	1088 &	1088 &	1102 &	1102 \\
16&	1349 &	1350 &	1349 &	1349 &	1344 &	1343 &	1139 &	1139 &	1139 &	1139 &	1139 &	1139 &	1139 &	1157 &	1157 \\
17&	1440 &	1442 &	1440 &	1440 &	1434 &	1433 &	1204 &	1204 &	1204 &	1204 &	1204 &	1204 &	1204 &	1222 &	1222 \\
18&	1538 &	1539 &	1538 &	1537 &	1530 &	1527 &	1258 &	1258 &	1258 &	1258 &	1258 &	1258 &	1258 &	1280 &	1280 \\
19&	1634 &	1635 &	1634 &	1633 &	1624 &	1621 &	1327 &	1327 &	1327 &	1327 &	1327 &	1327 &	1327 &	1349 &	1349 \\
20&	1737 &	1739 &	1737 &	1737 &	1727 &	1724 &	1392 &	1392 &	1392 &	1392 &	1392 &	1392 &	1392 &	1417 &	1417 \\
21&	1842 &	1843 &	1842 &	1841 &	1833 &	1830 &	1479 &	1479 &	1479 &	1479 &	1479 &	1479 &	1479 &	1504 &	1504 \\
22&	1959 &	1959 &	1958 &	1958 &	1953 &	1951 &	1569 &	1569 &	1569 &	1569 &	1569 &	1569 &	1569 &	1597 &	1597 \\
23&	2135 &	2136 &	2134 &	2134 &	2131 &	2128 &	1689 &	1689 &	1689 &	1689 &	1689 &	1689 &	1689 &	1717 &	1717 \\
24&	2359 &	2360 &	2358 &	2358 &	2352 &	2350 &	1801 &	1800 &	1800 &	1800 &	1800 &	1800 &	1800 &	1832 &	1832 \\
\hline
  &	1067 &	1068 &	1067 &	1066 &	1063 &	1062 &	911 &	911 &	911 &	911 &	911 &	911 &	911 &	923 &	923 
\end{tabular}
\end{small}
\end{center}
\end{table}

\begin{figure}[tbp]
\begin{center}
\includegraphics[width=120mm,angle=0,clip=]{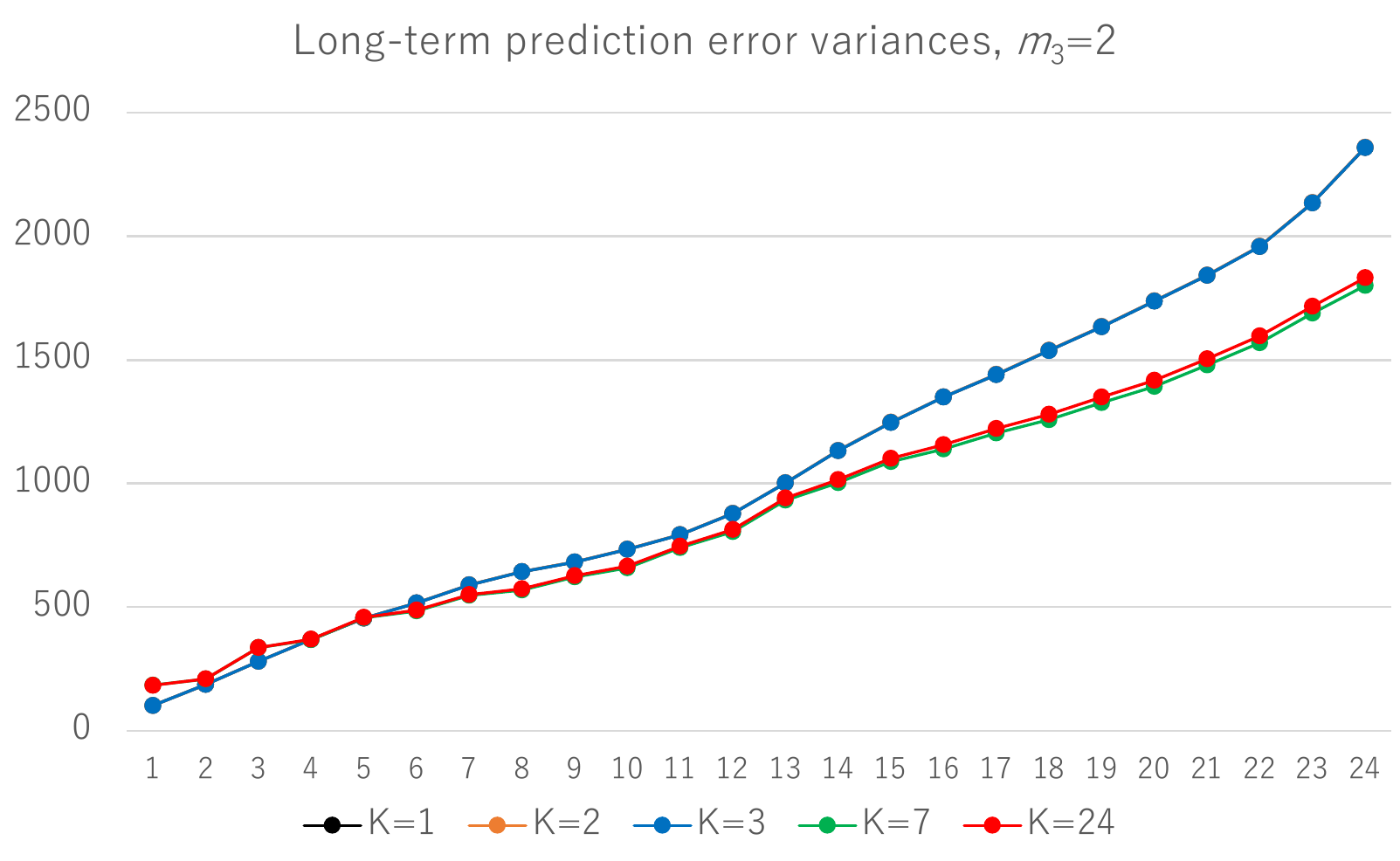}
\end{center}
\caption{The long-term prediction varainces of blsallfood data for increasing prediction horaizon (i=1,\ldots 20)
by the seasonal adjustment model with $m_1=2$, $m_2=1$ and $m_3=2$. 
$p$=1, 2 and 3.  }
\label{Fig_pred_error_variances_TSAR_M1=2}
\end{figure}

Figure \ref{Fig_pred_error_variances_TSAR_M1=2} show the long-term prediction error variances
for $p=1,2,3,7$ and 24.
The curves for $p =1$, 2 and 3 are visually indistinguishable.
The curve for $p=24$ is slightly larger than that obtained by $p=7$.

\begin{figure}[tbp]
\begin{center}
\includegraphics[width=130mm,angle=0,clip=]{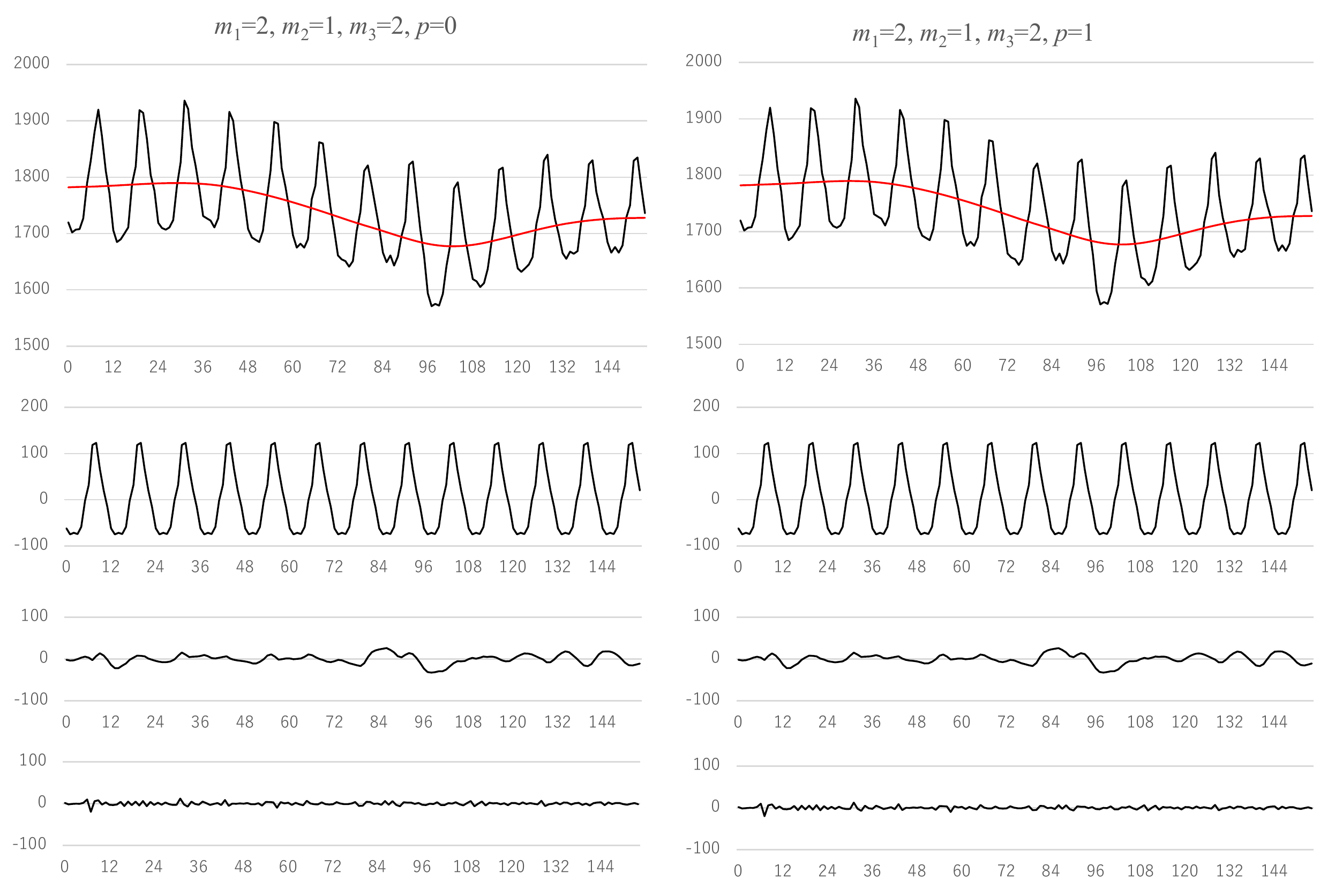}
\end{center}
\caption{The seasonal adjustment with AR component $m_1=2$, $m_2=1$ and $m_3=2$. 
Prediction lead time $p$=1 and 2. 
Top plot shows the data (black) and the mean of the trend(red), the second plot the seasonal component, the third plot the AR component and the botomn plot shows the noise component. }
\label{Fig_TSAR model with order M3=2_1}
\end{figure}

Figure \ref{Fig_TSAR model with order M3=2_1} shows the decomposition of the time series
into the trend, sesonal component, the AR component and the observation noise
obtained by $p=1$ (left plot) and $p=2$ (right plot), respectively.
This suggests that by selecting a value $p$ larger than 1,
it is very likely to obtain a model that has high increasing horizon prediction performance.
The results are indistinguishable.

\begin{figure}[tbp]
\begin{center}
\includegraphics[width=130mm,angle=0,clip=]{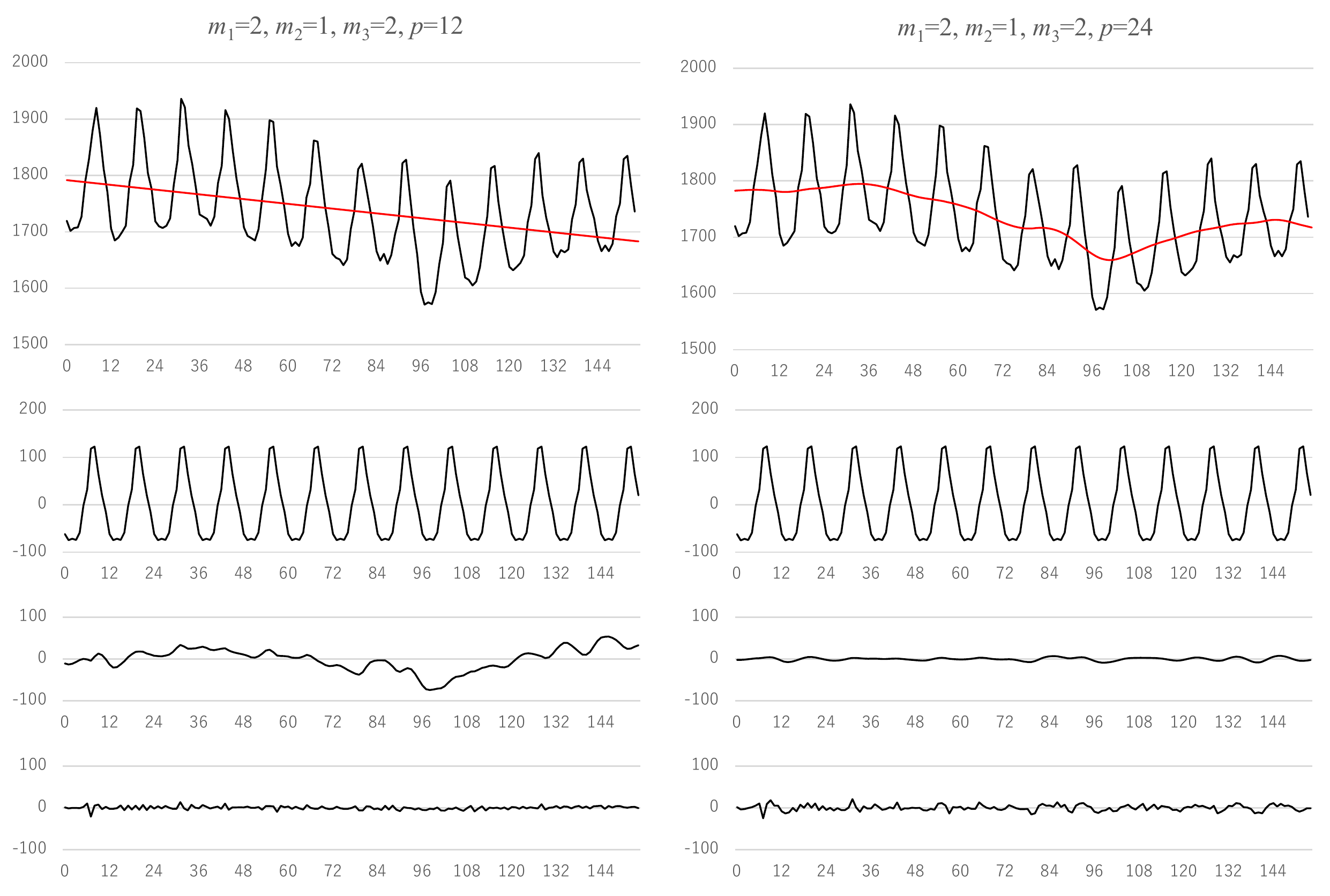}
\end{center}
\caption{The seasonal adjustment with AR component $m_1=2$, $m_2=1$ and $m_3=2$. 
Prediction lead time $p$=6 and 12. 
Top plot shows the data (black) and the mean of the trend(red), the second plot the seasonal component, the third plot the AR component and the botomn plot shows the noise component. }
\label{Fig_TSAR model with order M3=2_2}
\end{figure}

On the other hand, Figure \ref{Fig_TSAR model with order M3=2_2} shows the cases obtained by
$p=6$ (left plot) and 12 (right plot).
For $p=6$, the trend became a straight line and instead the AR component contains
a drift.
For $p=12$, the trend is slightly more variable than the trend by $p=1$ or 2 
and the AR component becomes very small.

From our expectation to the trend component, these estimate obtained by $p=6$ is a bit odd.
However, this estiamte has high increasing-horizon prediction performance.

\section{Concluding Remarks}
By the three examples, it can be seen that by specifying the prediction horizon p of the modified log-likelihood larger than 1, we can get a good long-term prediction performance.
The third example suggests that the seasonal adjustment model with AR component has resonable long-term prediction performance even with $p=1$.
This is probablly because the AR component can adapt to the local variation and increse the short-term prediction performance, but it does not diteriorate the long-term prediction because the prediction by stationary AR model converges to zero for large lead time.

\vspace{15mm}
\noindent{\Large\bf Aknowledgements}

This work was supported in part by JSPS KAKENHI Grant Number 18H03210.

\newpage

\noindent{\large\bf Appendix: Seasonal Adjustment Model with the First Order Trend Model}

\begin{table}[h]
\begin{center}
\caption{Long-term prediction error variances of the standard seasonal adjustment model with
$m_1=1$, $m_2=1$.}\label{Tab_Prdictive_variance_TSmodel_M1=1}

\vspace{2mm}
\tabcolsep=1mm
\begin{small}
\begin{tabular}{c|cccccccccccccccccc}
 &&&&&&&& $p$ \\
$j$  &1 &2 &3 &4 &5 &6 &8 &10 &12 &14 &16 &18 &20 &22 &24 \\
\hline
1 &250 &250 &250 &250 &250 &250 &250 &250 &250 &264 &254 &250 &250 &250 &250 \\
2 &268 &268 &268 &268 &268 &268 &269 &268 &268 &288 &275 &268 &268 &268 &268 \\
3 &347 &347 &347 &347 &347 &347 &348 &347 &347 &365 &354 &347 &347 &347 &347 \\
4 &415 &415 &415 &415 &415 &415 &416 &415 &415 &431 &422 &415 &415 &415 &415 \\
5 &477 &477 &477 &477 &477 &477 &478 &477 &477 &490 &482 &477 &477 &477 &477 \\
6 &518 &518 &518 &518 &518 &518 &519 &518 &518 &531 &523 &518 &518 &518 &518 \\
7 &561 &561 &561 &561 &561 &561 &561 &561 &561 &571 &564 &561 &561 &561 &561 \\
8 &597 &597 &597 &597 &597 &597 &597 &597 &597 &604 &598 &597 &597 &597 &597 \\
9 &649 &649 &649 &649 &649 &649 &649 &649 &649 &660 &652 &649 &649 &649 &649 \\
10 &686 &686 &686 &686 &686 &686 &687 &686 &686 &700 &691 &686 &686 &686 &686 \\
11 &714 &714 &714 &714 &714 &714 &715 &714 &714 &731 &721 &714 &714 &714 &714 \\
12 &743 &744 &744 &744 &744 &744 &745 &744 &744 &760 &750 &744 &744 &744 &744 \\
13 &984 &984 &984 &984 &984 &984 &981 &984 &984 &968 &973 &984 &984 &984 &984 \\
14 &1012 &1012 &1012 &1012 &1012 &1012 &1010 &1012 &1012 &1000 &1003 &1012 &1012 &1012 &1012 \\
15 &1073 &1073 &1073 &1073 &1073 &1073 &1071 &1073 &1073 &1065 &1067 &1073 &1073 &1073 &1073 \\
16 &1124 &1124 &1124 &1124 &1124 &1124 &1123 &1124 &1124 &1124 &1121 &1124 &1124 &1124 &1124 \\
17 &1163 &1163 &1163 &1163 &1163 &1163 &1163 &1163 &1163 &1173 &1166 &1163 &1163 &1163 &1163 \\
18 &1218 &1218 &1218 &1218 &1218 &1218 &1219 &1218 &1218 &1235 &1223 &1218 &1218 &1218 &1218 \\
19 &1272 &1272 &1272 &1272 &1272 &1272 &1273 &1272 &1272 &1292 &1278 &1272 &1272 &1272 &1272 \\
20 &1343 &1343 &1343 &1343 &1343 &1343 &1344 &1343 &1343 &1365 &1350 &1343 &1343 &1343 &1343 \\
21 &1379 &1379 &1379 &1379 &1379 &1379 &1382 &1379 &1379 &1415 &1393 &1379 &1379 &1379 &1379 \\
22 &1455 &1455 &1455 &1455 &1455 &1455 &1459 &1455 &1455 &1496 &1472 &1455 &1455 &1455 &1455 \\
23 &1527 &1527 &1527 &1527 &1527 &1527 &1531 &1527 &1527 &1570 &1546 &1527 &1527 &1527 &1527 \\
24 &1601 &1601 &1601 &1601 &1601 &1601 &1604 &1601 &1601 &1637 &1616 &1601 &1601 &1601 &1601 \\
\hline
    &891& 891&  891&  891&  891&  891&  891&  891&  891&  906&  896&  891&  891 & 891  & 891 
\end{tabular}
\end{small}
\end{center}
\end{table}

\begin{figure}[tbp]
\begin{center}
\includegraphics[width=100mm,angle=0,clip=]{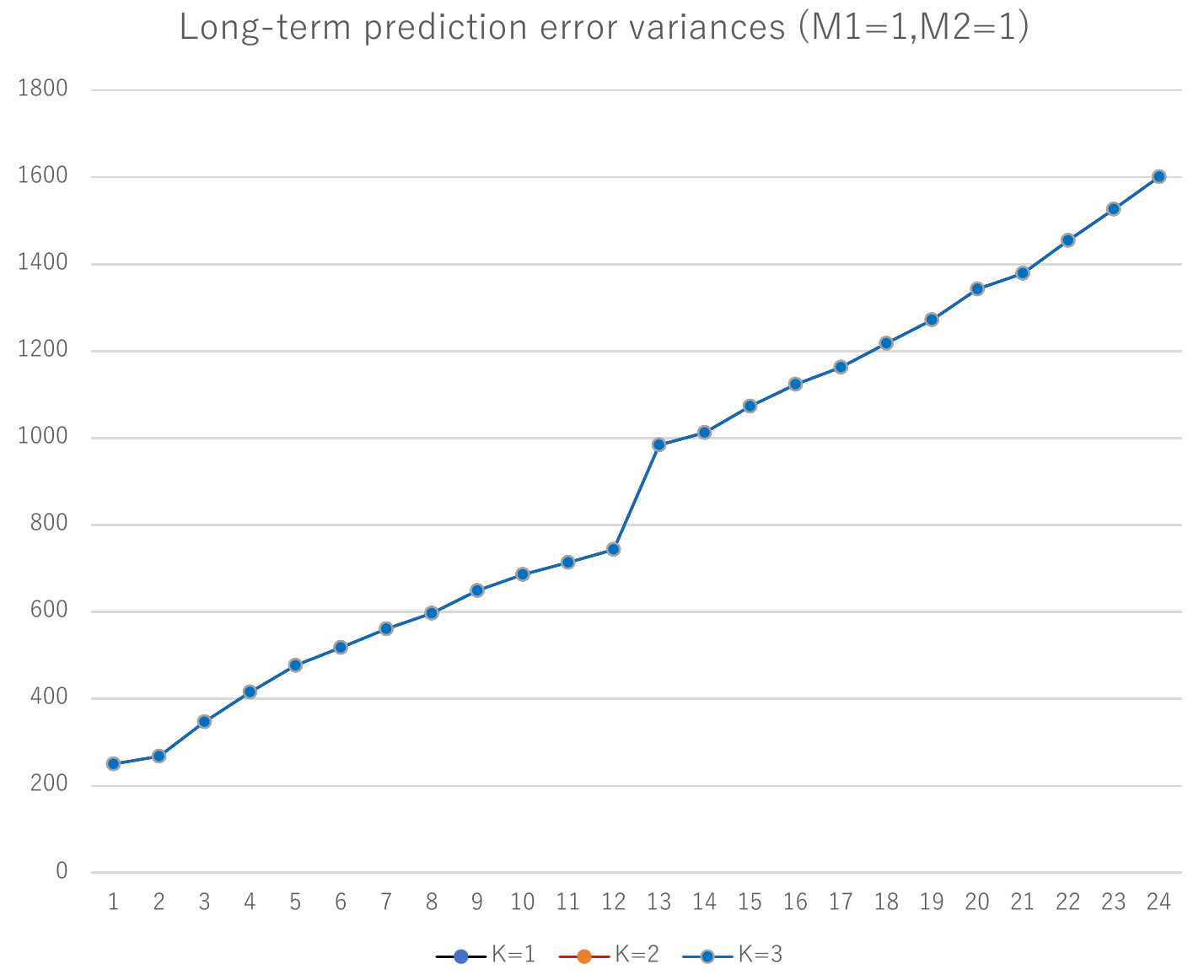}
\end{center}
\caption{The long-term prediction varainces of blsallfood data for increasing prediction horaizon (i=1,\ldots 20)
by the seasonal adjustment model with $m_1=1$ and $m_2=1$. 
NPRED=1, 2 and 3.  }
\label{Fig_pred_error_variances_TS_M1=1}
\end{figure}

\end{document}